\documentclass[a4paper,12pt]{article}
\pdfoutput=1
\usepackage{mathtext}
\usepackage{misccorr}
\usepackage{indentfirst}
\usepackage{graphicx}
\usepackage{amsmath}
\usepackage{amsbsy}
\usepackage{textcomp}
\usepackage{amssymb}
\usepackage{upgreek}
\usepackage{rotating}
\usepackage{cite}
\usepackage{lineno}
\newcommand {\eorp}{electron/positron}
\newcommand {\ep}{electron-positron}
\newcommand {\eps}{electrons/positrons}
\newcommand {\besh}{beamstrahlung}
\newcommand {\brsh}{bremsstrahlung}
\newcommand{\rBs} {radiative Bhabha scattering}
\newcommand{\isr} {initial state radiation}
\newcommand{\BDS} {beam delivery system}
\begin{document}
\title{Backgrounds at future linear colliders}
{\sf
\author{Oleg Markin\footnote{e-mail: markin@itep.ru}\\
{\small\emph{Institute for Theoretical and Experimental Physics, Moscow}}}
\date{\vspace{-5ex}}
\maketitle
\begin{abstract} A brief review of the background  for experiments at future $e^{+}e^{-}$ linear colliders is done.  Two sources  of background are discussed: the beam delivery system and  the interaction point.  The abundance of background muons, neutrons, photons and $e^{+}e^{-}$ pairs is quoted for different sub-detectors and both background sources. The background caused by the \besh\, is described  in more detail. Space distributions are sketched  and  the impact on  calorimeters is discussed for the background neutrons.
   
\end{abstract}
}
\maketitle
\section{Introduction}
The background  expected in experiments at future  $e^{+}e^{-}$ linear colliders substantially differs from that was at the LEP because of higher energy and much more narrow beam. Hits caused by background particles seriously complicate  the event reconstruction. The negative influence of the backgrounds on energy reconstruction comprises both worse pattern recognition and pile-up. Therefore the presence of the background should be carefully taken into account throughout the computer simulation of benchmark processes. After all, the load of detectors by background  signals determines the detector design as well as the choice of technologies for the detecting  systems.\footnote{This review represents studies of background completed to the autumn of  2012.}
\section{Backgrounds from the \BDS}
For the purpose of collimating both $e^+$ and  $e^-$ beams of the International  Linear Collider (ILC), three metallic primary collimators  will be situated at each side from the Interaction Point (IP). There are also protective masks in the beam tunnel. These elements  serve to purify the beam, but as they are placed rather close to IP, the  interaction of beam particles with these collimators and masks gives rise to background. The first of the collimators has an aperture $8\sigma_x\times65\sigma_y$, where $\sigma_x$ and $\sigma_y$ are the standard deviations of the transverse density of beam. That collimator is placed about 1,500~m from IP, while  two other collimators are about 1,300~m and 1,000~m from IP, respectively. 

\begin{figure}
\centering
\includegraphics[width=.8\textwidth]{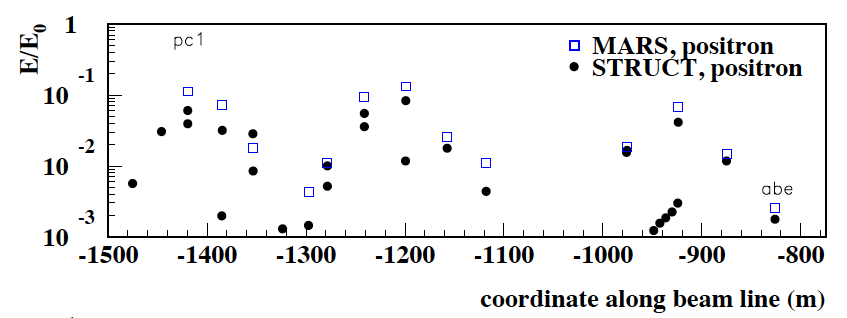}
\caption{ \textsf {Predictions of two different MC models for the loss of beam energy  along the beam line \cite{Drozdin_07}.}}
\label{loss}
\end{figure}

The presence  of the collimators leads to the loss of beam particles, this is shown  in fig.~\ref{loss} for positrons as predicted by two different MC codes. The magnitude of the loss is about 0.1\,\%; the lost positrons/electrons are being converted to  $\gamma$s, electron-positron pairs, muon-antimuon pairs and neutrons. These (background) particles contribute to so-called the Beam Delivery System (BDS) background. Besides, the BDS background includes photons originated from  bremsstrahlung at gas molecules in the beam tunnel and from the synchrotron radiation. In addition to the BDS background, there is an IP-related  background.   

\begin{figure}
\centering
\includegraphics[width=.9\textwidth]{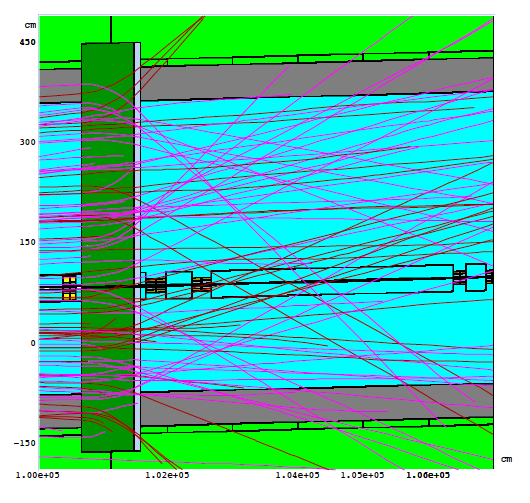}
\caption{ \textsf {Muon tracks in the spoiler region \cite{Denisov_06}.}}
\label{spoiler}
\end{figure}

\begin{figure}
\centering
\includegraphics[width=.8\textwidth]{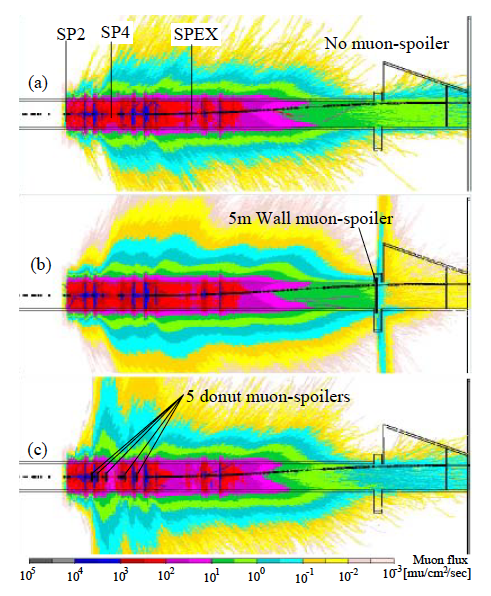}
\caption{ \textsf {Two-dimensional distributions of total muon flux
for (a) no, (b) wall- and (c) donut-shape muon spoilers \cite{Drozdin_07}. The notations SP2, SP2 and SPEX stand for the three primary collimators discussed in the text.}}
\label{sp_effect}
\end{figure}

The BDS neutrons can be created in photo-nuclear reactions.   
Muons can be produced in electromagnetic showers initiated by beam particles. 
The probability of the muon production is proportional to the squared charge of nuclei involved in the process, and amounts about $4\cdot10^{-4}$ \cite{Burkhardt_99}. Thus, the process results in about 4,000 muons for the 0.1\,\% loss of $10^{10}$ positrons/electrons per bunch-crossing (BX). A good way to minimize the number of those muons  is to sweep them away by the magnetic field created by iron spoilers  built purposely, cf. fig.~\ref{spoiler}.  There are two possible locations and   different constructions of the spoilers: wall- and donut-shape; fig.~\ref{sp_effect} demonstrates the effect of the two constructions. 

\begin{figure}
\centering
\includegraphics[width=.8\textwidth]{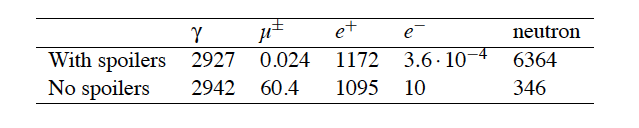}
\caption{ \textsf {Average number of particles per bunch at the beginning of detector from positron tunnel \cite{Denisov_06}.}}
\label{num_of_part}
\end{figure}

The spoilers drastically reduce the muon density, making it several thousand times lower at the entrance to the experimental hall. A side effect  of the spoilers is increasing of the number of neutrons, see table in the fig.~\ref{num_of_part}. The neutrons from the BDS background are not a problem since a concrete wall will be  placed at the entrance to the experimental hall. The vertex  system (VTX) is the only sub-detector not affected by the muon spoilers.

\begin{figure}
\centering
\includegraphics[width=.49\textwidth]{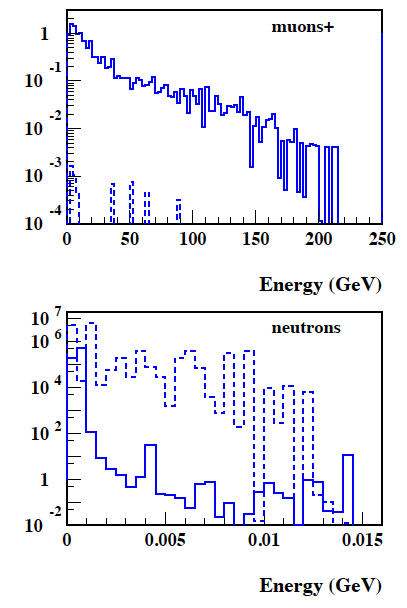}
\includegraphics[width=.49\textwidth]{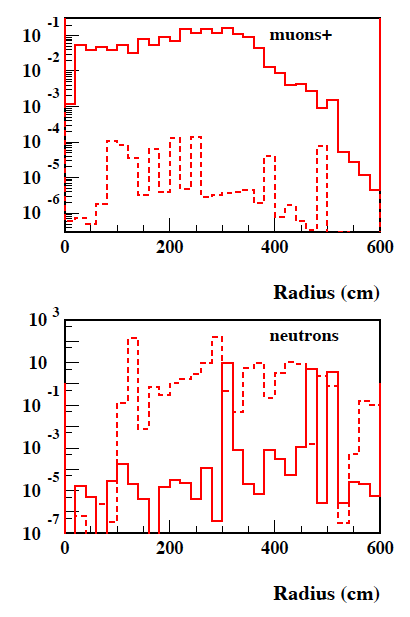}
\caption{ \textsf {Left: energy spectra of particles at the  detector (per bunch); solid line --- no spoilers, dashed line --- tunnel with spoilers. Right: radial distributions of particles at the detector  \cite{Denisov_06}. Shown are particles coming from positron tunnel only.}}
\label{distr}
\end{figure}

The energy distribution of the BDS muons and neutrons is shown in fig.~\ref{distr}. The BDS muons and neutrons have rather flat radial distribution within three meters from the beam axis, see fig.~\ref{distr}. The BDS photons and positrons/electrons are concentrated near or inside the beam pipe. The BDS background dominates in the muon detector while in the tracker and the VTX it gives the number of hits that does not exceed the IP background\cite{Denisov_06}. 

\section{Backgrounds related to the interaction point}

\subsection{Hard photons}

\begin{figure}
\centering
\includegraphics[width=.9\textwidth]{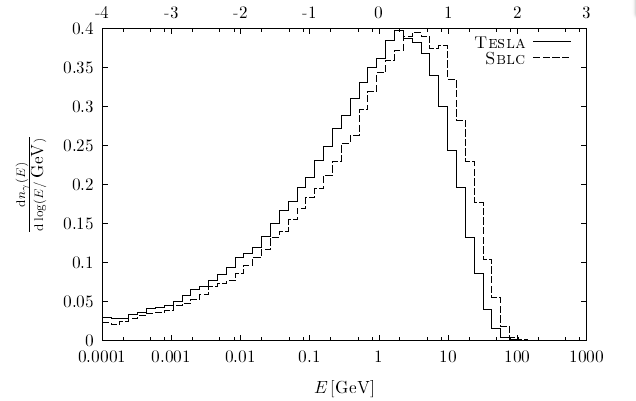}
\caption{ \textsf {The spectrum of \besh~photons for two ILC-like colliders \cite{Schulte_96}.}}
\label{besh_spectrum}
\end{figure}

At high-energy colliders the electromagnetic field of each bunch causes focusing of bunches of the opposite beam. That leads to  bending of \eorp~trajectories near the IP, which results in emission of hard photons. At the ILC, this process (called \besh) will give about two photons per a beam particle,  with the average energy of  \besh~photons being several percents of the beam energy \cite{Schulte_96}. The spectrum of the \besh~photons is shown in fig.~\ref{besh_spectrum}.  The \besh~photons are strongly focused in forward direction and, therefore, do not significantly contribute to the detector background. 
However, the \besh~photons  produce \ep~pairs that do contribute to the detector background either directly or through their backscatters. The photons can also produce neutrons when they hit components of the \BDS, such as collimators and the \besh~dump. 

\begin{figure}
\centering
\includegraphics[width=.6\textwidth]{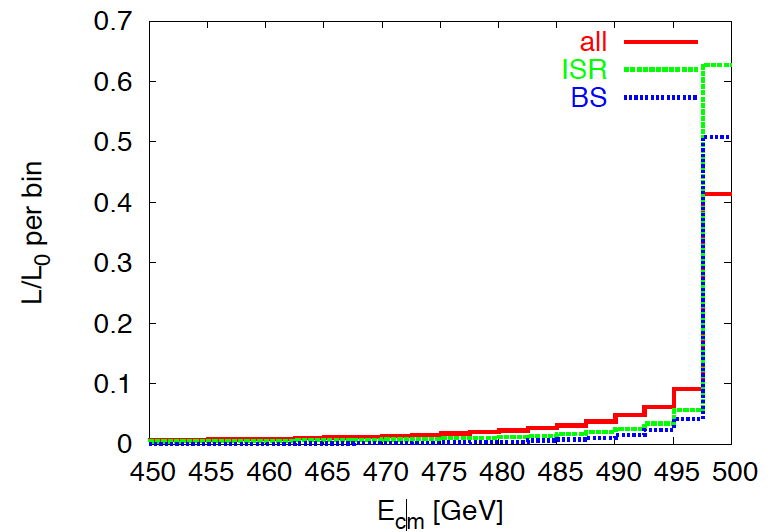}
\caption{ \textsf {Relative luminosity spectrum, considering beamstrahlung (BS), initial state
radiation (ISR) and both (all) \cite{Schulte_lec}.}}
\label{luminosity}
\end{figure}

\begin{figure}
\centering
\includegraphics[width=.6\textwidth]{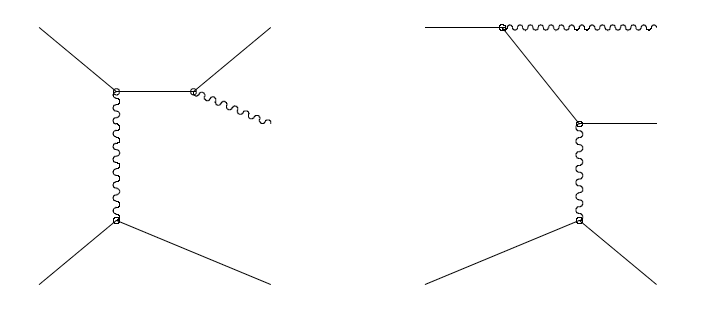}
\caption{ \textsf {Diagrams of \rBs.}}
\label{rbs}
\end{figure}

Besides, hard photons can be created in the ordinary initial state radiation. Both  the \isr~and the \besh~reduce the luminosity of colliders in the region close to  nominal energy, cf. fig.~\ref{luminosity}. A prominent source of hard photons is the collision of two beam particles resulting in \brsh, called the radiative Bhabha scattering, see fig.~\ref{rbs}. The photons created in the \rBs~are also focused in the forward direction, and escape through the beam pipe, but the remnant \eps~contribute to the detector background.

\subsection{Soft \eps}

The basic process that gives rise to the detector background is the creation of \ep~pairs additional to beam particles. An \ep~pair can be created by a \besh~photon in  the strong electromagnetic field of beams. This is referred  to as the coherent pair production. Under conditions of the Compact Linear Collider (CLIC),  the newly born pairs constitute several percents of beam particles. 

\begin{figure}[h]
\centering
\includegraphics[width=.8\textwidth]{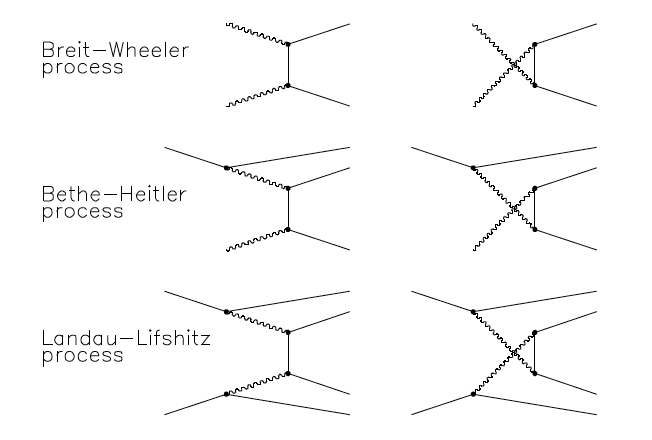}
\caption{ \textsf {Diagrams of \besh~pairs creation.}}
\label{pairs}
\end{figure}

\begin{figure}
\centering
\includegraphics[width=.5\textwidth]{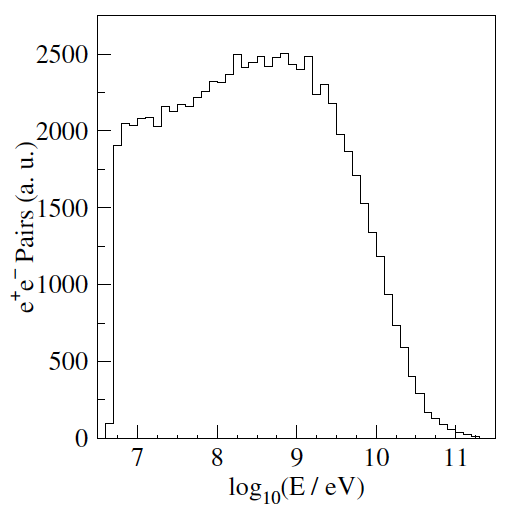}
\caption{ \textsf {The energy spectrum of electron-positron pairs simulated with GUINEA-PIG. The lower edge at 5~MeV corresponds to a momentum cut-off in the simulation \cite{Vogel_08}.}}
\label{pair_spectrum}
\end{figure}

At the ILC, another way of pair creation dominates, which is referred to as the incoherent pair production, when new \ep~pairs are created  as a result of interaction  of just two particles. Those can be either two real \besh~photons or a single real \besh~photon plus an \eorp, or  a couple of \eps, cf. fig.~\ref{pairs}. At the ILC with $\sqrt{s}=$0.5~TeV, there should be about 76,000 pairs per BX, with the average  energy 2.5~GeV per an \eorp~\cite{Maruyama_05}. The energy spectrum of the pairs is shown in fig.~\ref{pair_spectrum}. The interaction of photons can also give a couple of quarks followed by hadronization into minijets. 

\begin{figure}
\centering
\includegraphics[width=.6\textwidth]{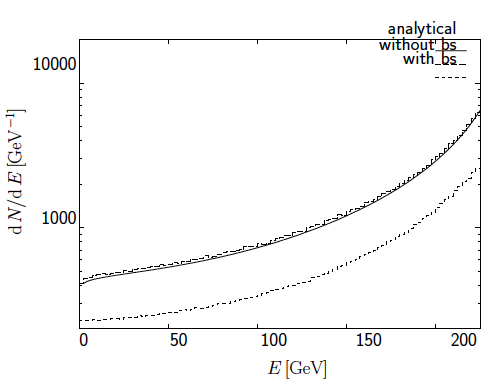}
\caption{ \textsf {Spectrum of \brsh~photons for an ILC-like collider. The letters 'bs' denote the so-called beam-size effect that almost halves the production rate. The similar reduction occurs for the \besh~pair production rate as well \cite{Schulte_96}.}}
\label{brsh_spectrum}
\end{figure}

An additional to \besh~source of off-energy \eps~is the radiative Bhabha scattering. For that process, the spectrum of remnant beam particles is rather flat at the low energy end \cite{Schulte_99}, see fig.~\ref{brsh_spectrum}. Under the  ILC-like conditions, after emission of a photon, the average energy of the scattered \eorp~is  equal to 50~GeV on average, and the number of corresponding electron-positron pairs is about $4\cdot 10^{4}$ per BX \cite{Wagner_01}.

The \eps~can be either focused or defocused by the beam field, which depends on the absolute value of particle momentum   and its  direction.  The remnant \eps~from \rBs~follow the appropriate beam and are thus focused. They will be dumped inside quadrupole magnets. The \eps~from  the \besh~pair production can follow  the positive beam direction as well as the negative one, which predetermines  focusing \cite{Schulte_99}. Most of them  initially have a small $\theta$-angle with respect to the beam axis. The final angle depends on the deflection by the beam field. The bigger transverse momentum $p_t$ they have, the smaller  $\theta$-angle they finally acquire. Therefore, only low-$p_t$ \eps~reach the detector where their range is limited by the main magnetic field, though. For the main part, those \eps~curl up and move longitudinally towards the quadrupoles. They may hit the forward detectors (LumiCal and BeamCal), the mask, the quadrupoles and the beam tube, producing secondaries, photons in that number,  which may convert to pairs inside sub-detectors. In the Time Projection Chamber (TPC) for example,  the \eps~from the pairs produced by the secondary photons are actually seen as lines (helices with tiny radius) parallel  to the field direction \cite{Vogel_08}. The number of such tracks is about 1,400 per BX, which also includes  backscatters from ECal and charged particles from the hadronic reactions in minijets.

\subsection{Neutrons}

A large amount of \eps~from pairs are deflected by the beam field and can hit the beam pipe and the quadrupole from inside. Thus there are two space origin of the IP-induced background: (i) directly the IP and (ii) backscatters from  the beam pipe itself and the very forward region, particularly the BeamCal. The latter serves  as an active absorber: it  provides shielding of the BDS elements from the incoherent pairs. 

\begin{figure}
\centering
\includegraphics[width=.7\textwidth]{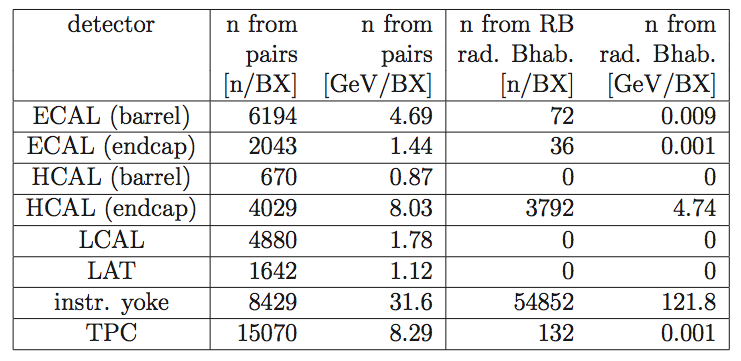}
\caption{ \textsf {The number of neutron hits and their energy deposited (except for TPC) in sub-detectors \cite{Wagner_01}.}}
\label{neut_det}
\end{figure}

Neutrons are produced by photons and \eps~that create electromagnetic showers where a photo-nuclear reaction may take place with a resonance at the photon energy about 10~GeV. For an ILC-like collider, the pairs created by \besh~photons produce about 70,000 neutrons per BX when they hit the mask and quadrupoles \cite{Wagner_01}. The remnant \eps~from \rBs~produce even more, about $3\cdot10^5$ neutrons per BX. Fortunately, they do not reach most of sub-detectors since they are produced at larger distance from IP, cf. the table in fig.~\ref{neut_det}. 

\begin{figure}
\centering
\includegraphics[width=.5\textwidth]{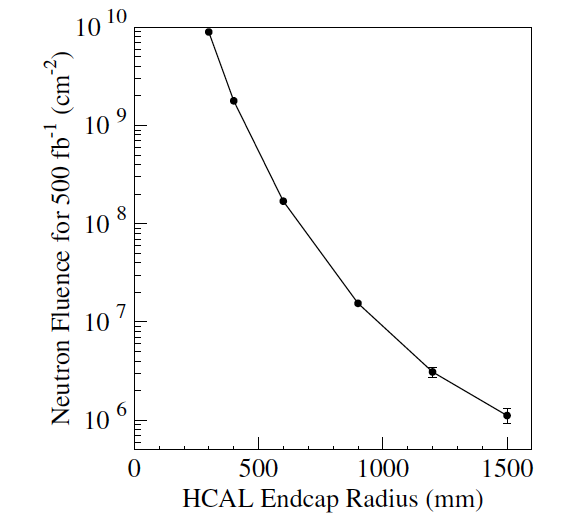}
\caption{ \textsf { The total neutron fluence at integrated luminosity 500~$fb^{-1}$ in the HCal endcaps, in dependency of the radial position \cite{Vogel_08}.}}
\label{neut_hcal}
\end{figure}

\begin{figure}
\centering
\includegraphics[width=.47\textwidth]{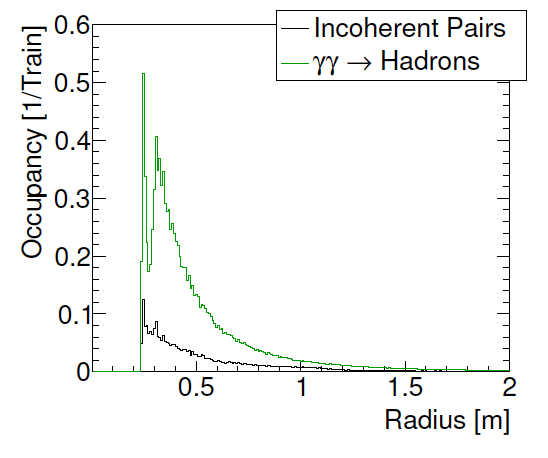}
\includegraphics[width=.49\textwidth]{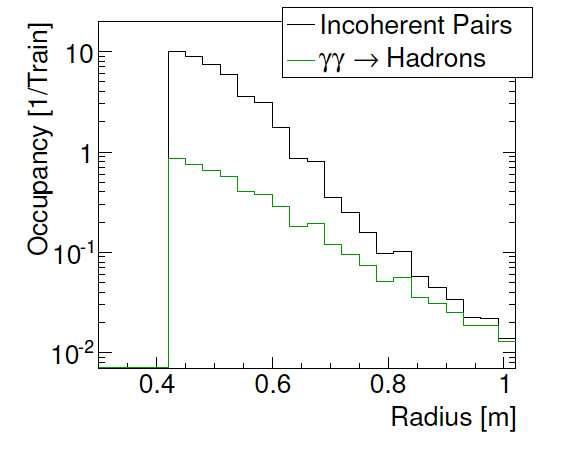}
\caption{ \textsf { The radial distribution of
the train occupancy per pad in ECal  (left) and per cell in HCal (right) endcap \cite{Dannheim_11}.}}
\label{CLIC_endcap}
\end{figure}

The neutrons created by the incoherent pairs at the BeamCal dominate the background in the endcaps of the Hadronic Calorimeter (HCal). The level of this background  is illustrated in  fig.~\ref{neut_hcal}. For the CLIC beam parameters (presumably at 3~GeV), the line corresponding to the background from incoherent pairs in the right plot in fig.~\ref{CLIC_endcap} actually represents  the neutrons. The cell occupancy in the HCal endcaps  at 40~cm from the beam axis is very high,  $\cal O$ $(10^{-2})$ hits per BX, steeply dropping outward. In the ECal endcaps, the dominating contribution to  cell occupancy is concerned with the minijet background. It decreases with a radial distance from the beam axis (i.e. with the polar angle of a final-state particle) not as fast as the neutron background in the HCal endcaps, cf. the left plot in  fig.~\ref{CLIC_endcap}. 

The signal of clusters created by particles from minijets can be rejected making use of precise time stamping. That has been studied throughout the preparation of the CLIC CDR by including the background from $\gamma\gamma \to $~hadrons  to simulation of several benchmark processes. For neutrons such a rejection procedure based on the time stamp might be less effective. Moreover, the substantial difference exists  between the neutron background and  the backgrounds from pairs or $\gamma\gamma \to $~hadrons  since  neutrons can deposit their energy  through  low-energy recoil protons in scintillator, which can produce ionization signals much exceeding those of MIP, amplifying thus the calorimeter response to the neutrons. Still, the background from neutrons created  by the incoherent pairs could have a significant impact on  the reconstruction of events only at very small polar angles because of the steep radial dependence, cf. fig.~\ref{CLIC_endcap}.   

Neutrons are also abundantly produced in water and concrete of both beam and \besh~dumps. Some amount of these neutrons return to the beam tunnel, but most of them are absorbed in the tunnel wall.  The rest, about 33,000 neutrons per BX, hit the detector yoke, but few of them reach sub-detectors \cite{Wagner_01}.

\end{document}